\documentclass[aps,prl,twocolumn,final,floatfix,superscriptaddress]{revtex4}
\usepackage{graphicx}
\usepackage{amssymb}
\usepackage{amsmath}

\begin{document}
\title{Velocity relaxation in a strongly coupled plasma}

	\author{G. Bannasch}
	\affiliation{Max Planck Institute for the Physics of Complex Systems, D-01187 Dresden, Germany}
	\author{J. Castro}
	\affiliation{Rice University, Department of Physics and Astronomy and Rice Quantum Institute, Houston, Texas, 77251, USA}
	\author{P. McQuillen}
	\affiliation{Rice University, Department of Physics and Astronomy and Rice Quantum Institute, Houston, Texas, 77251, USA}
	\author{T. Pohl}
	\affiliation{Max Planck Institute for the Physics of Complex Systems, D-01187 Dresden, Germany}
	\author{T.C. Killian}
	\affiliation{Rice University, Department of Physics and Astronomy and Rice Quantum Institute, Houston, Texas, 77251, USA}

\begin{abstract}
  Collisional relaxation of Coulomb systems is studied in the strongly coupled regime. We
  use an optical pump-probe approach to manipulate and monitor the dynamics of ions in an
  ultracold neutral plasma, which allows  direct measurement of relaxation rates in a regime
  where common Landau-Spitzer theory breaks down.
  Numerical simulations confirm the experimental results and display non-Markovian dynamics at early times.
\end{abstract}
\maketitle

More than half a century ago,  Landau \cite{lan36} and Spitzer \cite{spi67} derived simple expressions for Coulomb collision rates that have become fundamental to modern plasma physics. Precise knowledge of collisional relaxation rates is essential for understanding plasmas of all varieties. It is fundamental to energy exchange in multi-species systems \cite{spi67} and determines transport properties, such as self-diffusion rates as well as thermal and electric conductivities \cite{bsa05}.
The underlying theory, however, breaks down in strongly coupled systems such as Jovian planet interiors \cite{hor91} and dense-plasma experiments \cite{amt04}, which display strong correlations between particles.
Here, we present the first direct measurement of thermalization rates in an unmagnetized, strongly coupled plasma. Exploiting the very low temperatures in ultracold neutral plasmas \cite{kpp07,kil07}, we realize strong coupling conditions at low enough densities to enable direct time-resolved measurements via optical manipulation and imaging. The observations are supported by numerical simulations that moreover highlight the importance of non-Markovian relaxation effects.

In weakly interacting systems, that are either very hot and/or very dilute, relaxation is dominated by binary small-angle scattering events of distant particles. Consequently, a test charge traversing a single-species plasma of temperature $T$ and density $\rho$, undergoes Brownian motion with a corresponding damping coefficient \cite{lan36,spi67}
\begin{equation}\label{eq1}
\gamma(v)=\frac{2 \pi e^4 \rho }{\sqrt{m (k_{\rm B} T/2)^{3}}} \mathcal{R}(v)\ln\Lambda\;.
\end{equation}
where the factor $\mathcal{R}(v)$ derives from the so-called Rosenbluth potential
\cite{rmj57} and $m$ denotes the mass of the test particle and the plasma charges. The
term $\ln\Lambda$, known as the Coulomb logarithm, is determined by an upper cutoff 
for possible impact parameters that ensures convergence of the relaxation rate. In the 
original Landau-Spitzer derivation it is set equal to the Debye screening length, beyond 
which interactions are collectively screened by the surrounding plasma charges.

Equation (\ref{eq1}) is applied to a wide range of plasmas, but it is only valid when
spatial correlations in the system are weak. The degree of particle correlations can be
characterized by the ratio of their average potential and thermal energy, as expressed by
the Coulomb coupling parameter
\begin{equation}
\Gamma = e^2/(k_{\rm B}T a),
\end{equation}
where $a=(4\pi \rho/3)^{-\frac{1}{3}}$ is the average distance between the
plasma charges. A plasma becomes strongly coupled when $\Gamma > 1$, i.e. when interactions start to dominate thermal motion. The Coulomb logarithm can be written in terms of the coupling parameter, $\Lambda\sim \Gamma^{-3/2}$, showing that $\ln\Lambda$ turns negative for $\Gamma\gtrsim1$, and the Landau-Spitzer rate (\ref{eq1}) becomes entirely nonsensical in the strongly coupled regime.

\begin{figure}
\includegraphics[width=.99\columnwidth]{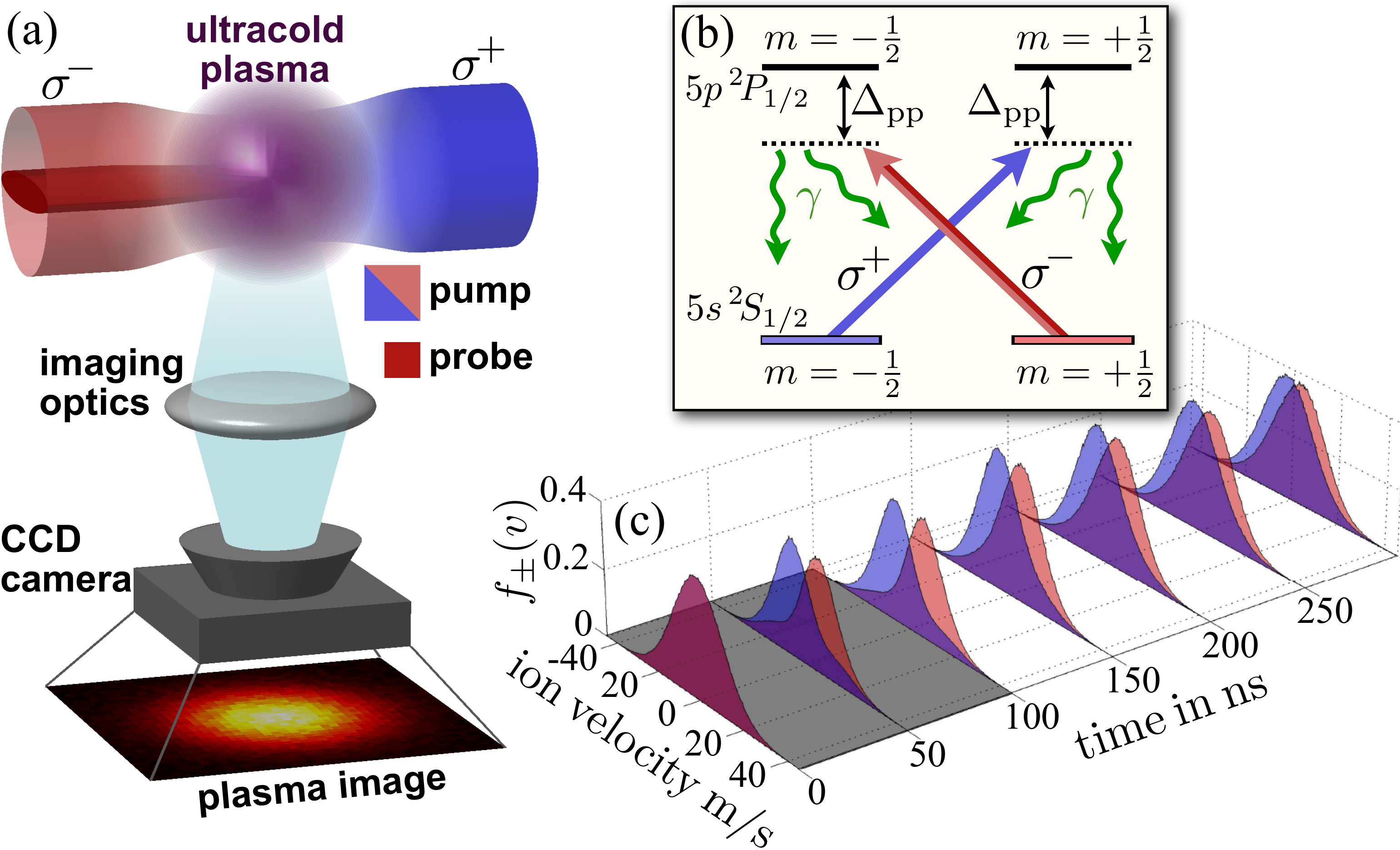}
\caption{\label{Fig1} 
  (color online).
  (a) Schematics of the experimental approach to probe
  ion relaxation in an ultracold plasma.  Two counter-propagating, circularly polarized
  lasers, detuned by $\Delta_{\rm pp}/2\pi=-20$\,MHz from the
  $\mathrm{5s\,^2S_{1/2}}-\mathrm{5p\,^2P_{1/2}}$ transition, optically pump population
  between the two ground-state magnetic sublevels ($m=\pm1/2$) of ions in an ultracold
  strontium plasma. The corresponding level-scheme is shown in (b), also indicating
  excited-state decay with the spontaneous emission rate $\gamma$. The optical pumping
  produces skewed velocity distributions $f_{\pm}(v)$ for each of the ground states, which
  we probe via ion fluorescence induced by a circularly polarized light sheet, applied at
  a variable time $t$ after optical pumping. A typical, simulated time evolution of the
  velocity distributions, $f_+$ (red) and $f_-$ (blue), during the optical pumping
  stage (gray) and subsequent relaxation is shown in (c).}
\end{figure}

Theoretical efforts to understand relaxation under strong-coupling conditions have largely focused on dense plasmas \cite{gms02,dda08,baa12,lem84,dwp98}, as produced by  intense-laser heating of solid-state samples \cite{scs98}. An accurate description of relaxation and transport processes is essential for the interpretation of these experiments \cite{rwm00} and, in particular, for optimizing conditions for inertial confinement fusion \cite{xh11}. Experimental probes of dense plasmas have advanced tremendously \cite{ggg08,knc08}, and allow indirect inference of relaxation rates through theoretical modeling of other observables \cite{glp07,agr08}. However, direct and precise measurement of relaxation rates remains an open challenge \cite{tsr06}, largely due to the fast dynamical time scales, complicated initial conditions and complex evolution at solid density.
 Ultracold neutral plasmas \cite{kpp07,kil07} present an appealing platform for studying strongly coupled plasma physics under simple and well-controllable conditions. Because of their low densities, ultracold neutral plasmas evolve slowly enough that the dynamics of many plasma parameters can be measured directly \cite{klk01,scg04,cdd05,flr07,bro08,mrk08,twr12,lgs07} with high temporal resolution.

We create an ultracold neutral plasma by photoionizing laser-cooled  strontium atoms in a
magneto-optical trap. 
Peak plasma density is varied from $\rho\sim10^{9}-10^{10}$cm$^{-3}$ by changing the delay between release of the trapped atoms and photoionization.
 By
tuning the ionization lasers, we  set the initial electron temperature to $T_{e}(0)=105$\,K
while the initial kinetic  energy of the ions equals the thermal  energy ($\sim10$mK) of the
laser-cooled atoms. This yields weakly  coupled electrons ($\Gamma_{\rm e}\sim10^{-2}$) and would
place the ions deep into the strongly coupled regime with   $\Gamma\sim10^3$.  However, photoionization produces  initially uncorrelated  ions, such that the subsequent  development of ion correlations results in strong heating \cite{bon96} to $T\sim1$K during  the first  few $100$ ns \cite{mur01,csl04,ppr05,bdl11}, and yields ionic Coulomb coupling parameters of order unity.
Ion temperatures are
determined by fitting the Doppler-broadened laser-induced-
fluorescence spectrum, while the
ion density is determined from absorption measurements of the plasma ions \cite{gls07}.

Due to the vast electron-ion temperature disparity, electronic screening of ion-ion interactions is considerably weaker than direct screening by the ions. For our conditions, the electronic Debye length is about $5$ to $10$ times larger than average particle distance $a$. The ionic component can thus be viewed as a classical one-component plasma, where the electrons provide a neutralizing background and electron-ion collisions play a negligible role. The latter drive the relaxation dynamics of dense plasmas \cite{gms02,dda08,baa12,lem84,dwp98,scs98,rwm00,xh11,ggg08,knc08,glp07,agr08,tsr06}. Here, the time scale for electron-ion relaxation greatly exceeds the duration of our measurement, which allows us to exclusively study ion-ion relaxation processes in the following.

To this end, we exploit the degeneracy of the Sr$^+$ $\mathrm{5s\,^2S_{1/2}}$
 ground state, i.e. the availability of two distinct electronic spin states (see Fig.\ \ref{Fig1}a). Initially, both spin states are equally populated, but we can
manipulate their populations by applying two counter-propagating laser pulses with
identical frequencies, detuned by $\Delta_{\rm pp}/2\pi=-20$\,MHz, opposite circular polarizations, and peak saturation parameters $s_0=6$ (see Fig.\ \ref{Fig1}a).
Due to the Doppler effect, this optically pumps population between the two spin states in a velocity-selective manner.
Ions with velocities around $v_z=\Delta_{\rm pp}/k$ along the wave vector ${\bf k}$ of the $\sigma^-$ laser
are transferred from $m=+1/2$ to $m=-1/2$ and vice versa around $-v_z$.
This produces skewed velocity distributions $f_{\pm}(v_z)$ for each spin state as shown in Fig.\ \ref{Fig1}c.
Note, that the state of the plasma and in particular the total velocity distribution $f(v_z)=f_{+}(v_z)+f_{-}(v_z)$ remains undisturbed and preserves its
Maxwellian form. Hence, the relaxation of $f_{\pm}$ is driven by an equilibrium
plasma with a well defined temperature $T$. In this way, our experiments realize the original
Landau-Spitzer construction of tagged test charges evolving in an equilibrium plasma background \cite{spi67}.

\begin{figure}
\includegraphics[width=.99\columnwidth]{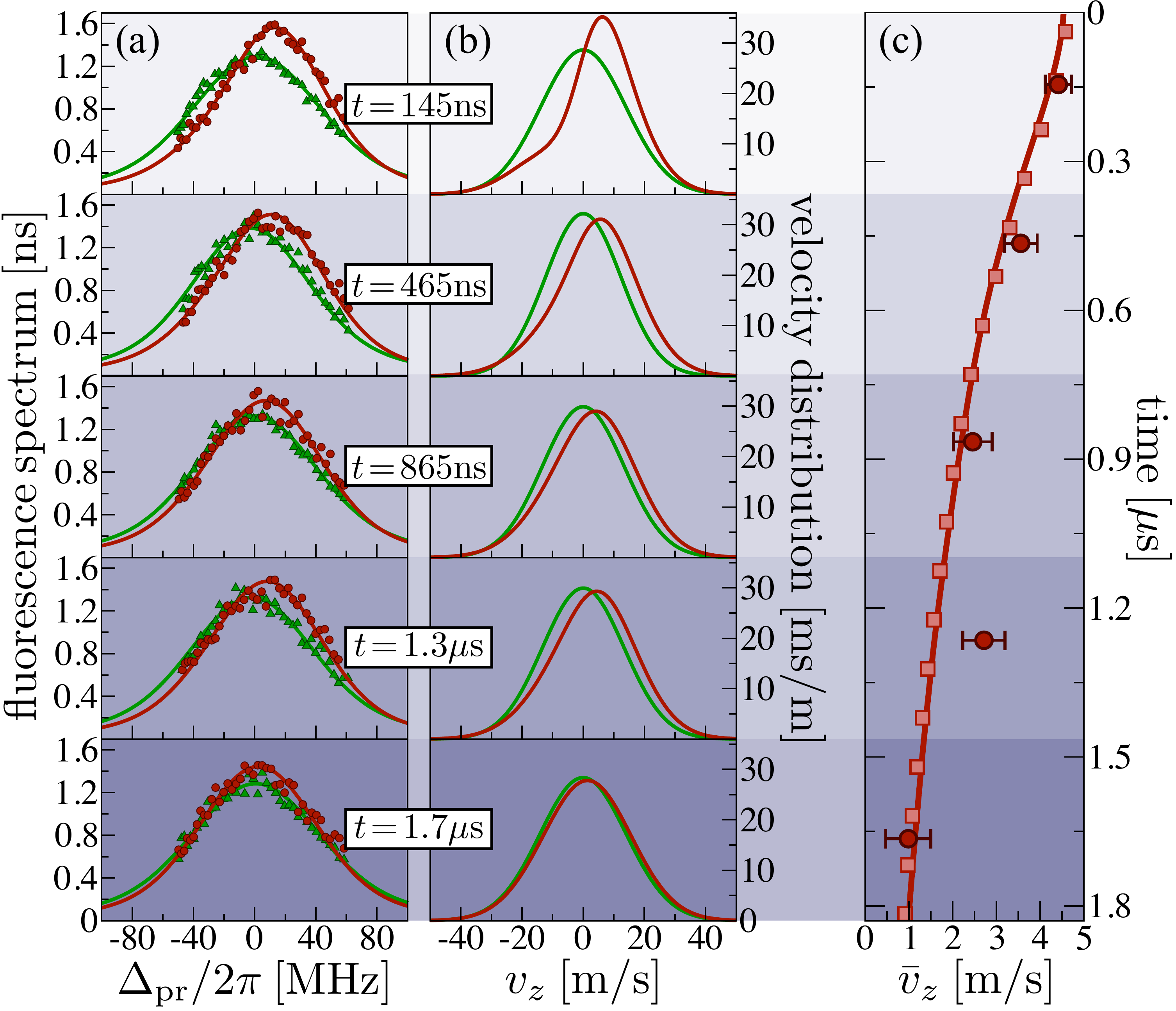}
\caption{\label{Fig2} (color online).  (a) Ion fluorescence spectra of a plasma with
  density of $\rho=10^9$cm$^{-3}$ and temperature of $T=2.2$K, corresponding to a Coulomb
  coupling parameter of $\Gamma=1.2$.  The circles show the measured spectra from the
  $m=+1/2$ states at different times $t$ after optical pumping. The spectra without
  optical pumping are shown by the triangles. They match the familiar Voigt profile
  arising from the underlying Maxwellian velocity distribution
  $f_{\rm eq}(v_z)=f_{+}(v_z)+f_{-}(v_z)$ of all plasma ions. The lines are fits of
  equation \eqref{eq:fit} to the experimental spectra, which yields the velocity
  distributions shown in (b). Panel (c) shows the corresponding average velocity (dots),
  along with the result of our quantum-classical simulations. The squares are obtained by
  calculating fluorescence spectra from the simulated velocity distributions, which are
  then analyzed just as the experimental spectra are.  The line shows the average velocity
  obtained directly from the simulations. The good agreement between both approaches
  confirms the accuracy of our experimental procedure for measuring the ion velocity.}
\end{figure}

At an adjustable time after optical pumping, we apply a third, less intense probe beam ($s_0=1.5$) with $\sigma^-$ polarization and an adjustable detuning $\Delta_{\rm pr}$ and record laser-induced fluorescence spectra perpendicular to the beam propagation (Fig.\ \ref{Fig1}a). We shape the probe beam into an ellipsoid that only interacts with a narrow central sheet of the plasma, and, thus, produces an image of a two-dimensional cut through the plasma cloud as shown in Fig.\ \ref{Fig1}a. We analyze fluorescence from a small central area where the density is nearly constant and the hydrodynamic expansion velocity \cite{lgs07} is negligible. The spectrum is given by a convolution of $f_+(v_z)$ and the Lorentzian profile of the probe transition, and can, thus, be used to determine the ionic velocity distribution.
Figure\ \ref{Fig2} shows a typical sequence of spectra obtained at different times after
optical pumping. Initially, one observes a considerable asymmetry with population
enhancement around $\Delta_{\rm pr}=-\Delta_{\rm pp}$ as expected. At later times, the spectrum approaches the familiar Voigt profile, indicating relaxation towards a Maxwellian velocity distribution within a few $100$ ns, which is close to the inverse of the corresponding plasma frequency, $\omega_{\rm p}^{-1}=\sqrt{m/4\pi e^2\rho}\approx200$ ns. The optical pumping and collisional processes maintain a simple form of the ion velocity distribution,
 \begin{equation}\label{eq:fit}
 f_{+}^{({\rm fit})}(v_z)=\frac{\alpha L(\delta_+,w)+1}{\alpha L(\delta_{+},w)+\alpha L(\delta_-,w)+2}f_{\rm eq}(v_z)\;,
 \end{equation}
from which we calculate the corresponding fluorescence spectrum and fit the measurements, using $\alpha$ and $w$ as free parameters. $L(\delta_\pm,w)\propto (w^2+\delta_\pm^2)^{-1}$ represents the Lorentzian lineshape of the optical pumping transition, with $\delta_{\pm}=kv_z\pm\Delta_{\rm pp}$. Equation (\ref{eq:fit}) follows from the steady state of the underlying optical Bloch equations, augmented by a simple Krook-type collision term \cite{bsa05}, and smoothly interpolates between the collisionless ($\alpha=0$) and the strong-collision ($\alpha\gg1$) regime \cite{cbm12}. As shown in Fig.\ \ref{Fig2}a, our data is well described by the fitted velocity
distributions, $f_+^{({\rm fit})}(v_z)$ (Fig.\ \ref{Fig2}b), from which we extract the average ion velocity
$\bar{v}_z=\int v_zf_+^{({\rm fit})}(v_z) {\rm d}v_z$ (Fig.\ \ref{Fig2}c).

We have also performed quantum-classical simulations of the laser-driven plasma dynamics
and subsequent relaxation in order to confirm our analysis procedure and to extend the
parameter range of our study. The calculations track the ion motion via classical
molecular dynamics (MD) simulations of the plasma ions, interacting by bare Coulomb interactions in a cubic simulation cell with periodic boundary conditions. The internal states of the ions are propagated according to the
 optical Bloch equations \cite{ae75} corresponding to the laser-driven $4$-level scheme shown in Fig.\ref{Fig1}b. This allows to follow the time evolution of the internal-state density matrix  during optical pumping alongside the phase-space
trajectory of each individual ion.  The quantum spin
evolution is coupled to the ion velocities via the Doppler shift, such that this approach yields the fully correlated internal and
translational plasma dynamics. The good agreement with our measurements (Fig.\
\ref{Fig2}c) demonstrates that this approach captures the essential physics of our
experiments. In addition, we used the
simulated distributions to generate theoretical spectra, which are then analyzed just as the
experimental spectra are. The excellent agreement between both results (Fig.\ref{Fig2}c) attests
to the accuracy of our experimental approach for extracting the average ion velocity based
on equation (\ref{eq:fit}).

\begin{figure}[t!] \centering
\includegraphics[width=.99\columnwidth]{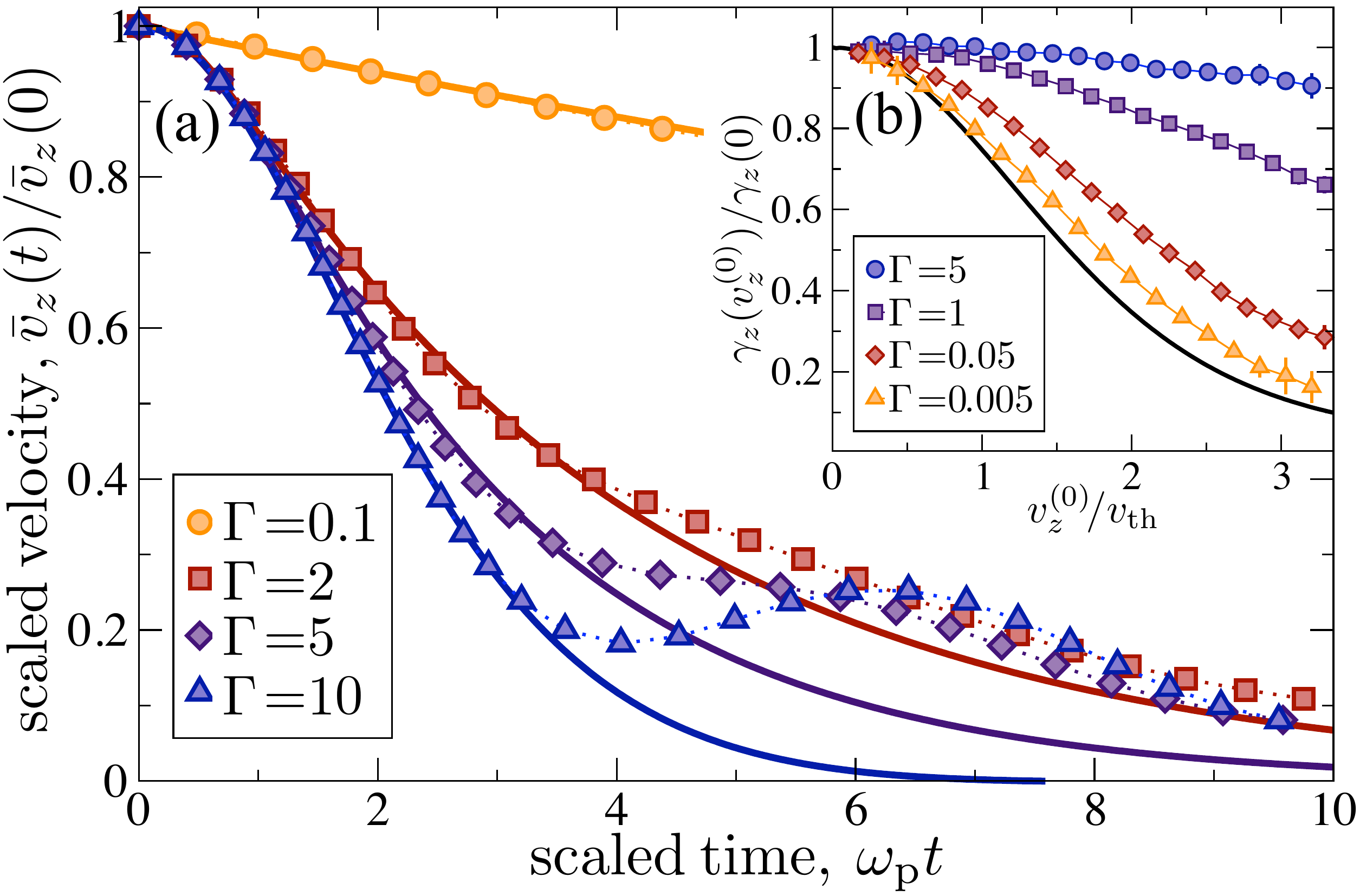}
\caption{
  (color online).
  (a) Simulation results for the  time evolution of the average
  ion velocity $\bar{v}_z$ (symbols), obtained for a narrow perturbation of the initial velocity
  distribution centered around $v_z^{(0)}=v_{\rm th}=\sqrt{k_{\rm B}T/m}$.
  Equations \eqref{eq:vmeanevolve} and \eqref{eq:kernel}, fitted to the data in the domain
  $t < 3 \, \omega_{\text{p}} t$ (lines), provide an excellent description of the short time
  dynamics.
  Panel (b) shows the extracted relaxation rate as a function of $v_z^{(0)}$. At small
  $\Gamma$ it displays a strong velocity dependence, which, however, weakens
  dramatically in the strongly coupled regime. The black line shows a comparison to the
  weak-coupling Landau-Spitzer prediction (see equation (\ref{eq1})) for $\Gamma=0.005$.}
  \label{Fig3}
\end{figure}

To determine the underlying relaxation rate, we model the relaxation dynamics by a non-Markovian damping term \cite{kub91}
\begin{equation}
\frac{d}{dt}\bar{v}_z (t) = -\int_0^t \mathcal{M}_z(v_z^{(0)}, t') \, \bar{v}_z(t-t^{\prime}) \,
  \text{d}t^{\prime}\;,
  \label{eq:vmeanevolve}
\end{equation}
where the stationary memory kernel $\mathcal{M}_z(v_z^{(0)}, t)$ accounts for retardation effects due
to the strongly coupled nature of the equilibrium plasma. Here, $\pm v_z^{(0)}$ are the
velocities on resonance with the pumping lasers, which show the strongest deviation from a
Maxwellian. The deviation is assumed to be well-localized in this derivation. The memory
time vanishes in weakly coupled plasmas, and
$\mathcal{M}_z(v_z, t)=2\gamma_z(v_z) \delta(t)$. This yields familiar exponential
relaxation with a damping constant
$\gamma_z(v_z)=\int {\rm d} {\bf v}_{\perp} \gamma(v) \, f_{\rm eq}({\bf v}_{\perp})$ that
coincides with the Landau-Spitzer result, averaged over the corresponding Maxwellian of
the transverse velocity ${\bf v}_{\perp}=(v_x,v_y)$.

However, temporal correlations become important for the strong coupling conditions of our experiments.
Following \cite{hmp75}, a simple Gaussian memory kernel
\begin{equation}
   \mathcal{M}_z(v_z^{(0)}, t) = \frac{2 \, \gamma_z(v_z^{(0)})}{\sqrt{2 \pi \tau^2}} \exp{\left(-\frac{t^2}{2\,\tau^2}\right)}\;,
 \label{eq:kernel}
\end{equation}
properly accounts for short-time correlations. The corresponding memory time $\tau$ is
connected to the average ion acceleration \cite{lev70} and can be obtained independently
from equilibrium simulations. For the time scales relevant to our experiments
($\omega_{\rm p} t \lesssim 3$), this simple theory is well confirmed by our simulations (Fig.\ \ref{Fig3}).

\begin{figure}[b!]
  \includegraphics[width=.99\columnwidth]{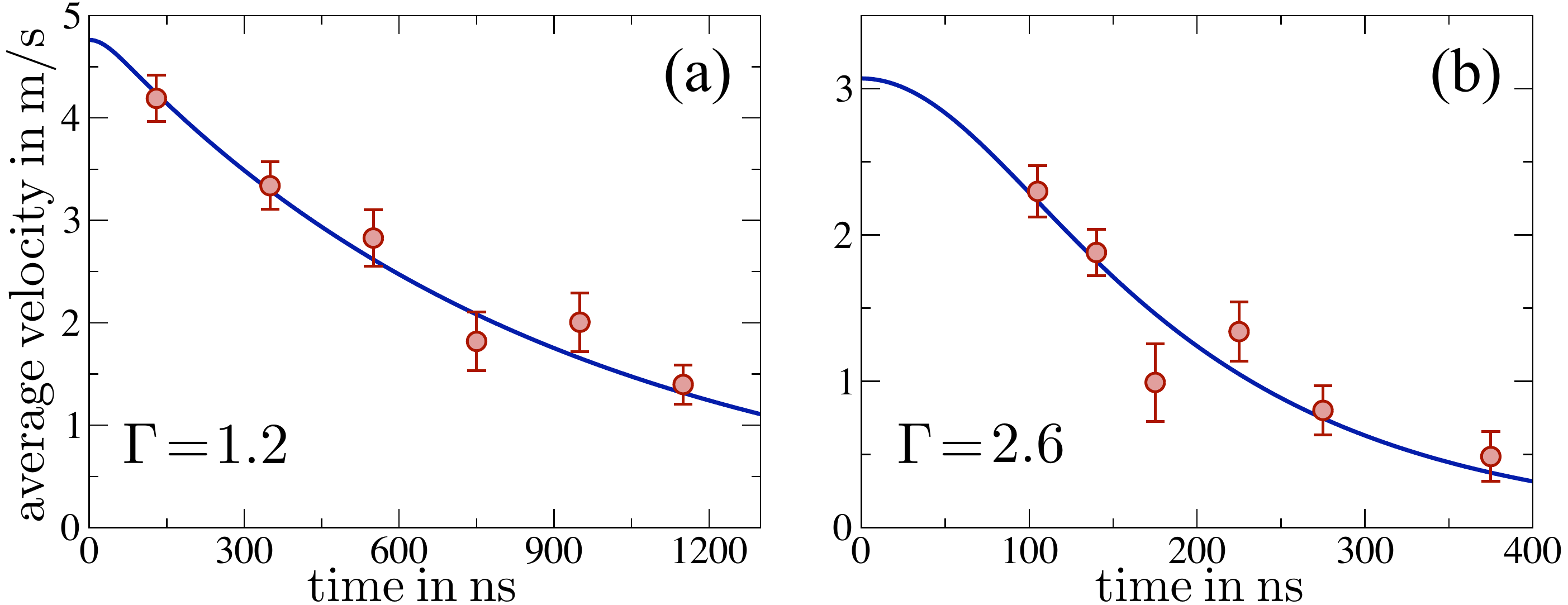}
  \caption{
    (color online).
    Measured time evolution of the average
    ion velocity $\bar{v}_z$ for two different sets of plasma parameters, (a)
    $\rho=3.1\cdot10^9$cm$^{-3}$, $T=3.3$K and (b) $\rho=4.4\cdot10^9$cm$^{-3}$, $T=1.7$K,
    corresponding to $\Gamma=1.2$ and $\Gamma=2.6$, respectively. The lines show a fit to
    the non-Markovian relaxation model, equations (\ref{eq:vmeanevolve}) and
    ($\ref{eq:kernel}$), from which we extract the corresponding relaxation rate $\bar{\gamma}$ shown in Fig.\ \ref{Fig5}.}
  \label{Fig4}
\end{figure}

An additional complication may arise from the velocity dependence of the relaxation rate (cf. equation (\ref{eq1})), which can make the dynamics of $\bar{v}_z$ depend on the specific form of $f_+(v_z)$.
To investigate this point we have performed simulations where the initial state transfer is done within a much narrower velocity range ($\Delta v_z= 0.05 \, v_{\rm th}$) centered around a velocity $\pm v_z^{(0)}$. Fitting eqs.(\ref{eq:vmeanevolve}) and (\ref{eq:kernel}) to our simulation results for different $v_z^{(0)}$ yields the velocity-dependent rates shown in Fig.\ \ref{Fig3}b. The velocity dependence weakens dramatically with increasing $\Gamma$ and nearly vanishes for our experimental conditions. In fact, the optical pumping affects almost the entire extent of the initially Gaussian velocity distribution (Fig.\ \ref{Fig2}b).
Since $\gamma(v)$ varies by less than $10\%$ over this range, we can apply equations (\ref{eq:vmeanevolve}) and (\ref{eq:kernel}) to our measurements and identify the extracted relaxation rate as the average $\bar{\gamma}\approx\int\gamma(v)f_{\rm eq}({\bf v}){\rm d}{\bf v}$. As Fig.\ \ref{Fig4} demonstrates, this simple approach provides an excellent description of our measurements.

\begin{figure}[t!]
  \centering
\includegraphics[width=.9\columnwidth]{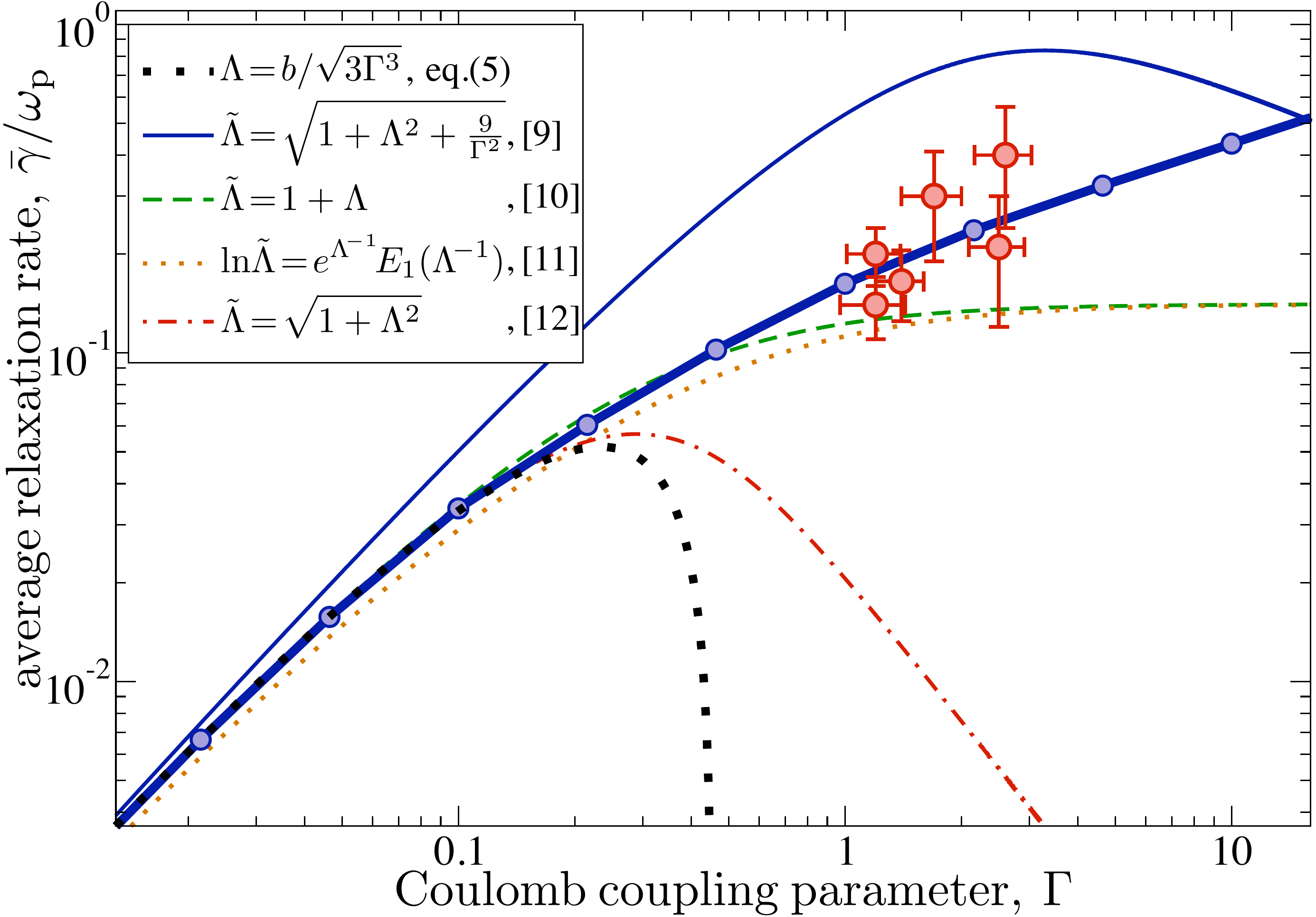}
\caption{ 
  (color online).
  Average relaxation rate as a function of coupling strength $\Gamma$. The
  large circles show the experimental results for different combinations of plasma density
  and temperature, within a range of $6\times10^8{\rm cm}^{-3}\lesssim\rho\lesssim5\times10^9{\rm cm}^{-3}$ and $0.8{\rm K}\lesssim T\lesssim2.8{\rm K}$. Using dimensionless quantities ($\bar{\gamma}/\omega_{\rm p}$ and
  $\Gamma$), the data collapses onto a universal curve, which verifies the expected
  Coulomb scaling and provides experimental evidence that, in the strongly coupled regime,
  the velocity dependence of the relaxation rate is negligible (see Fig.\ref{Fig3}b)
  within our measurement accuracy. The thick solid line is obtained from MD simulations for
  a wider range of Coulomb coupling parameters. In the weak-coupling limit it approaches
  the Landau-Spitzer form, equation (\ref{eq:clog}), shown by the thick dotted line. The
  other lines show different proposed extensions \cite{gms02,dda08,baa12,lem84} into the
  strongly coupled regime, obtained by replacing $\Lambda$ by $\tilde{\Lambda}$ in
  equation (\ref{eq:clog}) according to the expressions given in the figure. The function
  $E_1(x)=\int_{x}^{\infty}\!\!\tfrac{e^{-t}}{t}{\rm d}t$ denotes the exponential
  integral.}
  \label{Fig5}
\end{figure}

Figure \ref{Fig5} summarizes our main results, showing the relaxation rate $\bar{\gamma}$ for a wide range of parameters.
Upon scaling the rate by the ionic plasma frequency $\omega_{\rm p}$ and expressing the plasma temperature and density in terms of $\Gamma$, all data  collapse onto a single universal curve.
In the weakly coupled regime, the numerical results approach the familiar Landau-Spitzer form
\begin{equation}
  \label{eq:clog}
  \bar{\gamma} = a \Gamma^{\frac{3}{2}} \omega_{\rm p}\ln\Lambda\;,\quad\Lambda=\frac{b}{\sqrt{3\Gamma^3}},
\end{equation}
of the relaxation rate, with $a = 0.46$ and $b = 0.53$. Well into  the strongly coupled regime,
where equation\ (\ref{eq:clog}) predicts negative relaxation rates, we find good agreement between
experimental and numerical results.
A slight increase with $\Gamma$ is  evident in the experimental results
and is well reproduced by our calculations. As $\Gamma$
increases, both approach the plasma frequency, which sets the typical
time scale for ionic motion.

Figure\ \ref{Fig5} also includes recently proposed theoretical expressions based on  effective Coulomb logarithms that extend the Landau-Spitzer formula into the strongly coupled regime.  We note, however, that the scaled rate will depend on the mass ratio of the species considered, and previous theory has focused on ion-electron thermalization with an eye towards dense plasma applications. Although our experimental accuracy appears sufficient to  discriminate between different models, the additional mass-dependence, presently, limits such comparisons to a qualitative level. Perturbative corrections due to finite mass ratios have been investigated recently \cite{bsi09}. Extensions of existing models to equal-mass systems can now be subject to stringent tests through measurements in ultracold neutral plasmas.

The described pump-probe technique makes a whole new class of experiments possible. Laser
heating and cooling \cite{ppr04} will greatly stretch the range of accessible Coulomb
coupling parameters and allow exploration of the transition from an ideal to a correlated
plasma, extending more deeply into strongly coupled fluid regime. With improved time
resolution,  our approach will provide experimental access to velocity autocorrelations
and self-diffusion coefficients \cite{dal12}, which determine dynamic structure factors
and various transport processes. 

 This work was supported by the United States National Science Foundation and Department
 of Energy Partnership in Basic Plasma Science and Engineering (PHY-1102516) and the Air
 Force Office of Scientific Research (FA9550-12-1-0267).

\end{document}